\documentstyle[12pt,psfig,epsfig,axodraw]{article}
\advance\textheight by \topskip
\begin{document}
\tolerance=100000
\thispagestyle{empty}
\setcounter{page}{0}

\def\cO#1{{\cal{O}}\left(#1\right)}
\newcommand{\be}{\begin{equation}}
\newcommand{\ee}{\end{equation}}
\newcommand{\br}{\begin{eqnarray}}
\newcommand{\er}{\end{eqnarray}}
\newcommand{\ba}{\begin{array}}
\newcommand{\ea}{\end{array}}
\newcommand{\bi}{\begin{itemize}}
\newcommand{\ei}{\end{itemize}}
\newcommand{\bn}{\begin{enumerate}}
\newcommand{\en}{\end{enumerate}}
\newcommand{\bc}{\begin{center}}
\newcommand{\ec}{\end{center}}
\newcommand{\ul}{\underline}
\newcommand{\ol}{\overline}
\newcommand{\ra}{\rightarrow}
\newcommand{\sm}{${\cal {SM}}$}
\newcommand{\as}{\alpha_s}
\newcommand{\aem}{\alpha_{em}}
\newcommand{\ycut}{y_{\mathrm{cut}}}
\newcommand{\susy}{{{SUSY}}}
\newcommand{\Dir}{\kern -6.4pt\Big{/}}
\newcommand{\Dirin}{\kern -10.4pt\Big{/}\kern 4.4pt}
\newcommand{\DDir}{\kern -10.6pt\Big{/}}
\newcommand{\DGir}{\kern -6.0pt\Big{/}}
\def\Ecm{\ifmmode{E_{\mathrm{cm}}}\else{$E_{\mathrm{cm}}$}\fi}
\def\gluino{\ifmmode{\mathaccent"7E g}\else{$\mathaccent"7E g$}\fi}
\def\photino{\ifmmode{\mathaccent"7E \gamma}\else{$\mathaccent"7E \gamma$}\fi}
\def\mgluino{\ifmmode{m_{\mathaccent"7E g}}
             \else{$m_{\mathaccent"7E g}$}\fi}
\def\taugluino{\ifmmode{\tau_{\mathaccent"7E g}}
             \else{$\tau_{\mathaccent"7E g}$}\fi}
\def\mphotino{\ifmmode{m_{\mathaccent"7E \gamma}}
             \else{$m_{\mathaccent"7E \gamma}$}\fi}
\def\ML{\ifmmode{{\mathaccent"7E M}_L}
             \else{${\mathaccent"7E M}_L$}\fi}
\def\MR{\ifmmode{{\mathaccent"7E M}_R}
             \else{${\mathaccent"7E M}_R$}\fi}
\def\lsim{\buildrel{\scriptscriptstyle <}\over{\scriptscriptstyle\sim}}
\def\gsim{\buildrel{\scriptscriptstyle >}\over{\scriptscriptstyle\sim}}
\def\jp #1 #2 #3 {{J.~Phys.} {#1} (#2) #3}
\def\pl #1 #2 #3 {{Phys.~Lett.} {#1} (#2) #3}
\def\np #1 #2 #3 {{Nucl.~Phys.} {#1} (#2) #3}
\def\zp #1 #2 #3 {{Z.~Phys.} {#1} (#2) #3}
\def\pr #1 #2 #3 {{Phys.~Rev.} {#1} (#2) #3}
\def\prep #1 #2 #3 {{Phys.~Rep.} {#1} (#2) #3}
\def\prl #1 #2 #3 {{Phys.~Rev.~Lett.} {#1} (#2) #3}
\def\mpl #1 #2 #3 {{Mod.~Phys.~Lett.} {#1} (#2) #3}
\def\rmp #1 #2 #3 {{Rev. Mod. Phys.} {#1} (#2) #3}
\def\sjnp #1 #2 #3 {{Sov. J. Nucl. Phys.} {#1} (#2) #3}
\def\cpc #1 #2 #3 {{Comp. Phys. Comm.} {#1} (#2) #3}
\def\xx #1 #2 #3 {{#1}, (#2) #3}
\def\NP(#1,#2,#3){Nucl.\ Phys.\ \issue(#1,#2,#3)}
\def\PL(#1,#2,#3){Phys.\ Lett.\ \issue(#1,#2,#3)}
\def\PRD(#1,#2,#3){Phys.\ Rev.\ D \issue(#1,#2,#3)}
\def\preprint{{preprint}}
\def\Ord{\lower .7ex\hbox{$\;\stackrel{\textstyle <}{\sim}\;$}}
\def\OOrd{\lower .7ex\hbox{$\;\stackrel{\textstyle >}{\sim}\;$}}
\def\MCH {$\tilde\chi_1^+$}
\def \CH{{\tilde\chi}^{\pm}}
\def \LSP{\tilde\chi_1^0}
\def \SNU{\tilde{\nu}}
\def \BARSNU{\tilde{\bar{\nu}}}
\def \MLSP{m_{{\tilde\chi_1}^0}}
\def \MCH{m_{{\tilde\chi}^{\pm}}}
\def \MCHMIN {\MCH^{min}}
\def \ET{\not\!\!{E_T}}
\def \LL{\tilde{l}_L}
\def \LR{\tilde{l}_R}
\def \MLL{m_{\tilde{l}_L}}
\def \MLR{m_{\tilde{l}_R}}
\def \MSNU{m_{\tilde{\nu}}}
\def \PROCESS{e^+e^- \rightarrow \tilde{\chi}^+ \tilde{\chi}^- \gamma}
\def \PI{{\pi^{\pm}}}
\def \DM{{\Delta{m}}}
\newcommand{\bQ}{\overline{Q}}
\newcommand{\ad}{\dot{\alpha }}
\newcommand{\bd}{\dot{\beta }}
\newcommand{\dd}{\dot{\delta }}
\def \CH{{\tilde\chi}^{\pm}}
\def \MCH{m_{{\tilde\chi}_1^{\pm}}}
\def \LSP{\tilde\chi_1^0}
\def \MUL{m_{\tilde{u}_L}}
\def \MUR{m_{\tilde{u}_R}}
\def \MDL{m_{\tilde{d}_L}}
\def \MDR{m_{\tilde{d}_R}}
\def \MSNU{m_{\tilde{\nu}}}
\def \MLL{m_{\tilde{l}_L}}
\def \MLR{m_{\tilde{l}_R}}
\def \mhf{m_{1/2}}
\def \MST{m_{\tilde t_1}}
\def \RPVC{\lambda'}
\def\tth{\tilde{t}\tilde{t}h}
\def\qqh{\tilde{q}_i \tilde{q}_i h}
\def\t1{\tilde t_1}
\def \pt{p{\!\!\!/}_T}  
%#############################
%#$\tan\beta \approx 3$ 
\def\lapp{\mathrel{\rlap{\raise.5ex\hbox{$<$}}
                    {\lower.5ex\hbox{$\sim$}}}}
\def\gapp{\mathrel{\rlap{\raise.5ex\hbox{$>$}}
                    {\lower.5ex\hbox{$\sim$}}}}
%#############################
%\vspace*{\fill}
%\vspace{-0.5in}
%\begin{center}
%{\Large DRAFT v 1.3}\\
%{\bf Beware of Typos and technical mistakes!}
%\end{center}
\begin{flushright}
%{\today}\\
%{hep-ph/yymmdd}
\end{flushright}
\begin{center}
{\Large \bf
Constraining top squark in R-parity violating $\susy$ model 
using existing Tevatron data.
}\\[1.00
cm]
\end{center}
\begin{center}
{\large Subhendu Chakrabarti, Monoranjan Guchait  
{and} N. K. Mondal}\\[0.3 cm]
{\it Department of High Energy Physics\\
Tata Institute of 
Fundamental Research\\ 
Homi Bhabha Road, Bombay-400005, India.}
\end{center}

\vspace{2.cm}

\begin{abstract}
{\noindent\normalsize 
Signal of lighter top squark has been looked for using Tevatron data 
in the di-electron plus di-jet 
channel. 
We find that the mass of the lighter top squark when it decays 
dominantly in the electron plus jet channel, 
can be ruled out up to 220 GeV at 95\% C.L. using di-electron data. 
In the framework of R-parity breaking SUSY model
we exclude relevant R-parity violating couplings for a range of top 
squark masses and other SUSY parameters. The bounds on R-parity violating 
couplings are very stringent for the parameter space where lighter
top squark turns out to be the next to lightest supersymmetric particle. 
}
\end{abstract}
\vspace{2cm}
\hskip1.0cm
PACS no: 11.30.pb, 14.60.Cd, 14.80.Ly
%\vspace*{\fill}
%\vskip1.0cm
%\noindent
%\vspace*{\fill}
\newpage

\section*{I.~Introduction}
\label{sec_intro}
The Minimal Supersymmetric Standard model(MSSM)~\cite{susy} so far 
is one of the most credible candidate for the beyond standard model(SM) 
physics.
There is no single evidence of
supersymmetric (SUSY) particles, however, no observation can rule it out
either. Therefore, hunting for SUSY in the next generation of
colliders at Fermilab and at LHC experiments is a very challenging  
programme.
At present, from the non observation of SUSY signals in the past
experiments, mainly at LEP ~\cite{lepbound} and
Tevatron ~\cite {xsusy}, masses of SUSY particles have been constrained.

In MSSM, there is a mixing between the scalar superpartners of the two 
chirality states of fermions, $\tilde f_L$ and $\tilde f_R$.
The extent of mixing of these two chiral states is controlled by the
off diagonal term $ m_f(A_f - \mu \tan\beta)$ in the
sfermion mass matrix. It is obvious that the 
sfermions which are superpartners of the massive fermions
will have larger mixing effect because of the explicit dependence on 
the corresponding fermion mass
$m_f$ i.e. sfermions of 
third generation receive a large splitting between two mass eigen states.
Thus
the two chirality states of top squarks $\tilde t_L, \tilde t_R$, 
has large mixing,resulting in large splitting between the two physical mass
states 
$\tilde t_1,\tilde t_2$(assume $\MST \lsim m_{\tilde t_2}$)
~\cite{stopmix}. Moreover,
because of the large Yukawa coupling, the soft SUSY masses($m_{\tilde t_L},
m_{\tilde t_R}$)
also receive a large correction via the
renormalisation group equation ~\cite{run2susy} which can push the lighter
mass eigenstates, $\tilde t_1$, even below the top quark mass.
Consequently, in a certain region of SUSY parameter space
it may turn out to be the next to lightest SUSY particle(NLSP),
the lightest neutralino
$\tilde\chi_1^0$ being the lightest SUSY particle (LSP). 
It is to be noted that in the canonical SUSY searches at
colliders the missing energy due to the presence of
$\LSP$ which is assumed to be stable and non interacting, plays
a very crucial role~\cite{run2susy}.

In hadron colliders $\tilde t_1$ can be produced copiously, 
since it is colored and comparatively lighter than the  
other sparticles. Therefore, in the context of SUSY searches at 
hadron colliders, top squark searches has received a special attention.
The search strategy of top squark depend very crucialy on its decay 
pattern. As for example, the loop induced flavour changing neutral 
current decay mode~\cite{hikasa},
\br
\tilde t_1 \ra c \tilde \chi_1^0
\label{loopdk}
\er
yields acoplanar jets and missing energy from top squark pair production.
At Tevatron,  data corresponding to RUN-I experiment has been analysed 
to find 
top squark signal in this channel. Negative results have constrained
lighter top squark mass, $m_{\tilde t_1}\gsim $119 GeV (102 GeV) for
$m_{\tilde\chi_1^0}=$40(50) GeV~\cite{tevbound}. However, if
kinematically accessible, the top squark decays dominantly 
into a lighter chargino($\MCH$) and b quark, 
\br
\tilde t_1 \ra b + \tilde \chi_1^+ .
\label{chidk}
\er
Because of the cascade decays of $\CH_1$ into neutralino and massless
fermions, $\CH_1 \ra \LSP f \bar f'$, the top squark pair production leads
final states consisting leptons and jets accompanying by 
missing transverse energy~\cite{sender}.
Beside these popular decay modes eq.\ref{loopdk} and \ref{chidk}, 
there are 
also other interesting decay channels which yield a variety of 
signals in the colliders.   
All those decay modes will be discussed
in the next section. 

It is found that top squark mass upto $\sim$ 170 GeV can be probed
in RUN-II experiment at Tevatron with integrated 
luminosity 2 fb$^{-1}$ per experiments for the entire region of 
SUSY parameter space where
$\tilde t_1$ state appears to be NLSP~\cite{matchev} leading the decay mode
(eq.\ref{loopdk}) with 100\% branching ratio. For the scenario, when 
$\MST$ is heavier than lighter chargino mass, 4 fb$^{-1}$  
luminosity is required for the same
discovery limit in the dilepton plus missing energy channel which is
heavily contaminated by top backgrounds. The upgraded RUN-II experiment 
which may deliver high 
luminosity $\sim$ 20 fb$^{-1}$ may extend this reach upto 
$\sim$ 220 GeV~\cite{matchev}.

However, top squark phenomenology shows up new features
in the framework of R-parity violating(RPV) SUSY models.
In the SUSY models R-parity conservation(RPC) is assumed to forbid the decay
of proton ensuring conservation of lepton and baryon numbers, L and B 
respectively. However,
one can avoid proton decay problem by invoking either L or B conservation.
Thus one can have two kind of RPV SUSY models corresponding to L or B violation.
As a consequence, in any SUSY cascade decay process within the framework
of RPV SUSY model, the LSP can have decay 
modes either in the leptonic 
or hadronic channels leading multileptons and multijets
in the final states with or without missing energy. The prospects of SUSY 
searches at Tevatron in the context of 
R-parity breaking SUSY model has been investigated in great detail
~\cite{run2rpv}.

In RPV SUSY model, the superpotential is 

\br
W  = W_{MSSM}+  W_{{R{\!\!\!/}_p}},
\er
where $W_{MSSM}$ is the superpotential containing yukawa type of
interactions giving masses to the fermions and $ W_{R{\!\!\!/}_p}$
corresponds to the potential containing terms which violate L and B
numbers,
\br 
W_{{R{\!\!\!/}_p}}=
\lambda_{ijk} L_i L_j E_K +
\lambda'_{ijk} L_i Q_j D_k + \lambda''_{ijk} U_i D_j
D_k + \mu_i L_i H_2 .
\label{rpveq}
\er
Here $i, j, k$ are the generation indices, 
$\lambda,~\lambda'$ and $ \lambda''$ are the dimensionless yukawa
couplings. The superfields L,Q represent the SU(2) doublets for leptons and 
quarks respectively where as singlets U,D and E stand for up type, 
down type quarks and charged leptons respectively.   
Last term in eq.~\ref{rpveq} mixes the mass terms of the 
lepton and higgs doublets ~\cite{rpv}.    
However, in the present case
we will work in the context of spontaneous~\cite{spon} RPV  
neglecting this bi-linear term in eq.~\ref{rpveq}.

In this work we focussed on the top squark decays in 
RPV SUSY model. Considering only lepton number violation 
the $\lambda'_{i3j}$ coupling which leads to a 
new decay channel of 
$\tilde t_1$,
\br
\tilde t_1 \ra \l + q . 
\label{rpvdk}
\er
As a consequence, in this scenario the pair of top squark 
production are signalled by dilepton plus di-jets. As we know, the
identical final states also appear due to the pair 
production of  
Leptoquark($LQ$) assumed to be the composite object of lepton and 
hadron~\cite{lepto} and its subsequent decays into the corresponding 
lepton and 
quark
\br
LQ \ra \ell + q .
\label{lepdk}
\er
At Tevatron experiments, the searches for all three
generations of $LQ$ which are signalled by ee + 2jets, $\mu\mu+$2 jets and
$\tau \tau+$ 2 jets respectively were carried 
out~\cite{cdflepto,d0lepto1,d0lepto2}.
This remarkable similarity between the final states due to the pair 
production 
of top squarks in RPV 
SUSY scenario and Leptoquark pair production and its subsequent decays 
via eq.\ref{lepdk} motivated us to exploit the existing data
corresponding to this Leptoquark searches to constrain  
top squark mass which is involved
in the production mechanism. Recall that $\t1$ 
states has 
also other RPC decay modes as well
which will be discussed in the next section. 
The branching ratio(BR) of $\t1$ in the RPV channel,  
eq.\ref{rpvdk}, is controlled by RPV couplings and as well as other 
SUSY parameters
which are involved in determining the decay rates corresponding to 
RPC decay channels.
In this work our main goal is to analyse top squark signal
in the di-electron plus di-jet channel and compare it with the data  
corresponding to this final state
which was used for 1st the generation of Leptoquark search at Tevatron.
This leads to lower limit of top squark mass for a
given BR of top squark in the RPV channel. Moreover, the
RPV couplings($\lambda'_{13j}$) can be excluded for a given top squark mass
and SUSY parameter space. In the case of dimuon channel, we found in the 
paper of Ref~\cite{d0lepto2} that the background corresponding to this 
channel has been analysed using neural network (NN) analysis and the 
same NN has been trained to analyse signal. 
Because of this we could not perform the analysis for this channel,
while the investigation of the ditau 
channel due to the $\lambda'_{33j}$ RPV coupling is now under 
progress. Hereafter, whenever we will refer to data, that will 
correspond to the data in the di-electron plus di-jet channel.  
Earlier there was also attempt to constrain RPV couplings using 
top quark data~\cite{asesh}.   

It may be recalled that these RPV decay channels drawn a lot of attention for 
the possible 
interpretation to explain the excess of high $Q^2$ events reported 
in H1 and ZEUS experiments few years back~\cite{Q2}. In that context,
implications of these channels was examined at Tevatron \cite{dp}.

It is worth mentioning here the existing bounds of the RPV couplings 
which are relevant for the present purpose.
From the direct production of $\t1$ in electron-proton 
collision at HERA, the H1 experiment predicts a bound, 
$\lambda'_{131} \lsim $ 0.05(.02) at 95\% C.L for $m_{\t1}$=200(100) 
GeV~\cite{H1}. A bound on $\lambda'_{131}$ also exist from 
atomic parity violation(APV), $\RPVC_{131} \lsim $0.07 
at 95\% C.L for $m_{\t1}$=200 GeV~\cite{apv}. The forward-backward 
asymmetry in $e^+e^-$ collision predicts the bound on  
$\lambda'_{132} \lsim
0.28$ for $m_{\tilde t_L}=$100 GeV~\cite{rpv132}.
The stringent bound on $\lambda'_{133}$ comes from neutrino data.  
In the framework of RPV SUSY model the neutrino masses can be generated 
from the tree level contributions due to the bi-linear terms or loop 
contributions from the trilinear $\lambda$ and $\lambda'$ couplings
~\cite{gb}. The detailed phenomenological analysis has been done 
using neutrino data including trilinear and bi-linear couplings
~\cite{abada1,abada2}. The most favoured large mixing angle solution constraints
the trilinear couplings  $\lambda'_{133} \lsim 10^{-4}$ assuming 
$M_{susy}=100$ GeV~\cite{abada2}.

We have organised our paper as follows. In section II, we discuss 
branching ratios of top squark decay into various channels 
in the context of RPV SUSY model. Our analysis is described in section III 
followed by our results in section IV. Finally we conclude with a summary 
in section V.

\section*{II.~Top squark decay in R-parity violating SUSY model}
\label{stopdk}
As mentioned in the last section that the lighter state of top
squark, $\tilde t_1$, has many phenomenologically interesting decay 
modes depending on
its mass. The most dominant decay mode of $\tilde t_1$, if kinematically
accessible, is via lighter chargino state, $\tilde\chi_1^\pm$ and $b$ quark,
(eq.\ref{chidk}).
In the absence of this two body charged current decay 
mode, the flavour changing neutral
current decay mode and the four body
decay channel, into a b quark, 
the LSP($\tilde \chi_1^0$) and two approximately massless 
fermions~\cite{boehm,djouadi}, 
\br
\tilde t_1 \ra b \tilde\chi_1^0 f \bar f'
\label{4body}
\er
are the only available decay modes and very competitive to each other.
This 4-body decay channel occurs via many diagrams involving
a variety of heavier SUSY particles 
in the intermediate state.
In the papers of ref.~\cite{boehm,djouadi},
the decay pattern in this channel 
has been discussed elaborately over a wide  
range of SUSY parameter space assuming R-parity conservation. 
Surprisingly, in certain region
of SUSY parameter space the 4-body decay mode takes over the
loop level decay mode and the branching ratio may shoot up to
$\sim$100\%. It implies that the signal corresponding to the
neutral current decay modes of top squarks will be suppressed.    
The impact of this 4 body decay modes
in the context of top squark searches at upgraded 
Tevatron has been discussed~\cite{yaan,spdas}.
Beside these decay channels of $\t1$, there are also a few other decay modes
which may be interesting from the phenomenological point of view. 
As for example, if kinematically
accessible, the three body decay mode to bottom quark and a $W$ boson
or a charged Higgs scalar $H^\pm$, and a neutralinos, $\tilde t_1 \ra
b W^\pm \tilde\chi_1^0$ or $\tilde t_1 \ra b H^+ \tilde\chi_1^0$ may
open up~\cite{porod1}. Moreover, in the light
slepton scenario which is viable in some SUGRA models~\cite{sugra} 
$\tilde t_1$ decays via the final states containing sleptons, 
$\tilde t_1 \ra b \tilde\ell \nu, b \ell \tilde\nu$~\cite{guchait}.

In RPV SUSY model, top squark decay channel, eq.~\ref{rpvdk} opens 
up due to the interaction  
\br
{\lambda'}_{i3j} \ell_i \tilde t_L q_j + h.c
\er
which is a subset of the Lagrangian given by eq.\ref{rpveq}.
The species of the lepton and quark depend
on the choice of $i$ and $j$ respectively, where $i,j$=1,2,3. 
Since we are not restricted to the jet flavour therefore $j$ can be 
of anything 1-3 in the decay process,eq.~\ref{rpvdk}. 
%\br
%\tilde t_1 & \ra & e + d; \nonumber \\
%           & \ra & \mu +  s;  \nonumber \\
%	   & \ra & \tau +  b; \ \ 
%\er

Neglecting the fermion masses, the decay width
of $\tilde t_1$ in the RPV channel (eq.~\ref{rpvdk}) is given by
\br
\Gamma_{R{\!\!\!/}}(\tilde t_1 \ra \ell + q)=
\frac{ {\lambda'}_{i3j}^2 \cos^2\theta_{\tilde t} }{16 \pi}
\MST^2
\label{rpvgama}
\er
$\theta_{\tilde t}$ is the mixing angle in the top squark sector;
it appears due to the replacements of $\tilde t_L$ by the
physical states $\t1$. 
The BR in this channel is
given by,
\br
\epsilon =\frac{\Gamma_{R{\!\!\!/}}(\tilde t_1 \ra l + q)}
{\Gamma_{R{\!\!\!/}}(\tilde t_1 \ra l + q) + \Gamma(\tilde t_1 \ra RPC)}
\label{epsi}
\er
where $\Gamma(\tilde t_1 \ra RPC)$ stands collectively for the total 
decay width in all accessible RPC decay modes of $\tilde t_1$ states.
The rate of those RPC decay modes depend on SUSY parameters, 
particularly on the SU(2) gaugino mass $M_2$ (assuming electroweak
gaugino mass relation, $M_1 \simeq \frac{M_2}{2}$, where $M_1$ is the
U(1) gaugino mass), the higgsino 
mass parameter $\mu$ and $\tan\beta$ - the ratio of the vacuum expectation values 
of two higgs doublets  
which give up type and down type quark masses. 
Beside these parameters, it involves  
other sparticle masses and parameters, e.g. mass of 
sleptons ($m_{\tilde\ell}$),
mass of squarks ($m_{\tilde q}$) and A-terms, the trilinear 
couplings.
We investigate the relative rates of the RPV decay mode over the RPC
decay modes for a
wide range of SUSY parameter space. We compute the decay 
widths corresponding to the loop induced flavour changing decay mode 
(eq.~\ref{loopdk}) and four body decay width(eq.~\ref{4body}) following    
the expressions given in the paper of Ref.~\cite{hikasa,boehm,djouadi}. 

In fig.1, we present contour plots for the constant BR ($\epsilon$ ) 
corresponding to
the RPV decay modes of $\t1$  
in the $\MST $
- $\lambda'_{i3j}$ plane for a fixed set of SUSY parameters: $M_2=125$ GeV
$\mu=400$ GeV, $\tan\beta=4$, $m_{\tilde q}=300$ GeV and 
$m_{\tilde\ell}=200$
GeV and $\cos\theta_{\tilde t}=0.8$. We find for low $\MST$, 
where the two body
chargino decay mode (eq.\ref{chidk}) is kinematically inaccessible
the RPV decay
modes has comparatively appreciable rates even
for very small value of $\lambda'_{i3j} (\sim 10^{-4})$. 
In contrast, for higher $\tilde t_1$ masses, the two body charged current
decay mode will open up 
resulting in the supression of the RPV decay. 
In this case the RPV mode will be important
for $\lambda'_{i3j} \sim$ O(0.1).

\section*{III.~Squark production: Event Analysis}

At Tevatron top squark pairs are produced via quark-antiquark 
annihilation and
gluon gluon fusion,
\br
q\bar q,~g g  \ra \tilde t_1 \tilde t_1^* .
\er
Since, this pair production mechanism is dominated by the QCD 
process, 
the cross section depends solely on $\MST$~\cite{kane}. 
The SUSY-QCD corrections enhance the
cross section by another $\sim$ 30\% over most of the SUSY parameter 
space ~\cite{spira}. We estimate the cross section setting 
renormalisation and 
factorisation scale at $Q^2 = \hat s$ and use CTEQ3L~\cite{cteq} for the 
parton distribution functions. The typical top squark pair 
production cross 
section ranges
from $\sim$ 10.-0.1 pb for $\MST \sim$100-200 GeV 
at $\sqrt{s}=1.8$ TeV.

We analyse the top squark pair production in the 
di-electron plus di-jet channel which originates because of the top 
squark decay via eq.~\ref{rpvdk}.
As we discussed earlier that the 
same type of event topology also appears 
due to the pair production of 1st generation of Leptoquark  
and its subsequent decay through eq.~\ref{lepdk}. 
The dominant SM 
backgrounds corresponding to this signal come from Drell-Yan process.
The other sources of backgrounds are due to the production of 
$WW$, $W+$jets, $t\bar t $ and $Z$+jets followed by the leptonic decay of
vector boson. 
For the present purpose we do 
not estimate the cross sections for each of these  
background processes. 
Instead we closely follow the analysis as described in the 
paper ~\cite{d0lepto1} for 1st 
generation of Leptoquark searches at D0 experiment, where the kinematic 
cuts are chosen to effectively suppress these backgrounds.

The number of signal events for a given $\MST$ is given by,
\br
n_{sig} = \sigma_{\tilde t_1 \tilde t_1^*}. \epsilon_{d}.
{\cal L}. \epsilon
\label{sigcs}
\er
Here, $\epsilon_{d}$ 
stands for the detection efficiency which includes the acceptance
efficiency due to the 
selection and as well as background rejection cuts and all 
systematic efficiencies like trigger and lepton identification 
efficiencies. 
The luminosity for the given 
set of data is given by ${\cal L}$ and $\sigma_{\tilde t_1 \tilde t_1}$ 
represents the top squark pair
production cross section and $\epsilon$ is the BR of $\tilde t_1$ 
decay into lepton plus jet.
We generate events using {\tt PYTHIA}(V6.206)
~\cite{pythia} producing top squark pair which are forced to   
decay via eq.~\ref{rpvdk} in order to assume the 100\% 
branching ratio of top squark in this RPV channel. 
Our intention is to compute the accepted efficiency for a given set of 
kinematic cuts corresponding 
to this di-lepton plus di-jet final states. 
We take into account the effect of 
initial and final state radiation as well as 
fragmentation effects in the event generation. We adopt the following 
strategy to compute the 
signal cross section for a given $\MST$ and then 
comparing our
results with the existing data we 
obtain limits of $\MST$ and RPV couplings.

${\bullet}$ First, for a given $\MST$, we compute acceptance 
efficiency for the signal process by generating events 
using {\tt PYTHIA}~\cite{pythia}  
applying the same set of 
cuts which are  used in the analysis for the first generation 
of Leptoquark searches
~\cite{d0lepto1}. Then we multiply the respective 
systematic efficiencies, e.g. trigger and lepton identification efficiencies
appropriately with the acceptance efficiency to obtain overall 
detection efficiency $\epsilon_d$.

$\bullet$ Secondly, from the existing data corresponding to this 
final state, for a given $\epsilon$ and $\MST$, we estimate the 
cross section limit using eq.\ref{sigcs} following Bayesian approach
with a flat prior probability distribution of cross section.
The systematic and statistical uncertainties  
are included taking Gaussian prior distribution of each of them.   
Then the limit of cross section is compared with the theoretical 
prediction for a given $\MST$ setting a fixed value of $\epsilon$.

$\bullet$ And finally, repeating this procedure for various $\MST$, 
the upper limit
of $\epsilon$ can be obtained for each $\MST$.
Eventually, this upper limit of $\epsilon$ can be translated to 
obtain upper bound of RPV coupling for a given set of SUSY parameters.

Now, in the following we discuss event analysis for the di-electron 
plus di-jet channel.

The pair production of top squark and subsequent decays of each top squarks
in the RPV channel via coupling 
$\lambda'_{13j}$ 
\br
\tilde t_1 \tilde t_1^\ast \ra e e + qq 
\er
results the final states, e~e + 2 jets.
As discussed earlier that it is identical to the final states
due to the pair production of 1st generation 
of $LQ$ and its subsequent decay via eq.~\ref{lepdk}. For our signal 
Monte Carlo, we 
followed the analysis very closely as described in Ref.~\cite{d0lepto1}
where search for 1st generation $LQ$ with same final states has been 
described 
and then compare our results with the data.
As we mentioned above that the dominant SM backgrounds 
corresponding to this di-electron
final state are from Drell-Yan production with two or more jets, 
$t \bar t$ production and multijet events in which two jets are 
miss identified as electrons{\footnote {The probability of faking a jet as
electron is very small $\sim $O($10^{-3})$~\cite{d0lepto1}}.
In our simulation we selected events in the hadronic and 
electromagnetic calorimeter cells in pseudorapidity and 
azimuthal angle($\phi$) of
size $\Delta \eta \times \Delta \phi=0.1 \times 0.1$.
%with energy deposition
Cells with $E_T > $ 1 GeV are taken as initial seeds to form 
calorimetric tower. 
For the jet reconstruction we use the routine {\tt PYCELL} in {\tt PYTHIA}
~\cite{pythia}. Jets are reconstructed with cone radius 0.7 and 
accepted only 
those which has transverse energy $E_T >$ 8 GeV and are  
smeared by 0.5$\times \sqrt {E_T}$. 
Events are subjected to the following sets of cuts mentioned in 
the paper of Ref.~\cite{d0lepto1} ,

\begin{enumerate}
\item
Two electrons  with $E_T^e >$ 20 GeV and within the
coverage for central calorimeter(CC) $|\eta| <$1.1 and for endcap 
calorimeter (EC) 1.5$<|\eta| <$2.5.
\item
At least two jets having $E_T^j >$ 15 GeV and $|\eta| <$2.5. 
\item
Isolation between electrons and jets are maintained by requiring   
$\Delta R_{ej} >$
0.7 where $\Delta R=\sqrt{\Delta\phi^2 + \Delta\eta^2}$
\item
Events of having di-electron invariant mass between 
82 $< M_{ee}<$ 100 GeV are rejected. 
\item
 The total visible transverse energy($S_T$) satisfy 
the cut $S_T >$ 350 GeV where $S_T = H_T^e + H_T^j$, 
$H_T^e=$ sum of the $E_T$ of the two electrons; $H_T$=sum of the
$E_T$ of all jets.
\end{enumerate}
The cuts 1-3 are the event selection cuts where as cut 4-5 are the 
background rejection
cuts. The cut 4 is used to avoid the contamination due to the events  
from Z production.  
%The total number of background events which is dominated by Drell-Yann
%process 
%is 101 which is in good agreement with data sample $92.8 \pm 13.8$
%events. 
In the Drell Yan process, leptons and jets are not so hard
as in the case of signal process for $\MST \sim$100-200 GeV.
So a cut on the sum of the transverse energies of visible particles in the 
final state drastically reduce this background. 
The cut 5 serves that purpose and brings
down the level of background to a negligible level.
We found that the signal acceptance efficiencies which is only due to the
kinematic cut effect vary from 0.1 - 15\% for
$\MST=$ 80-200 GeV. 
We take into account the electron identification efficiencies which are 
$74 \pm 3$\%, $66 \pm 4$\% and $68 \pm 9$\% for CC-CC, CC-EC and EC-EC 
regions respectively~\cite{d0lepto1} by multiplying appropriately 
with the acceptance 
efficiencies which are obtained from {\tt PYTHIA},
and we refer this as a detection efficiency. 
In table.1, we show the detection efficiencies folding 
all other systematic 
efficiencies together for various choices of $\MST$. For
a given $\MST$, we can obtain from the knowledge of detection efficiency 
the number of di-electron events for a given $\epsilon$ and luminosity. 
The total integrated luminosity is $123 \pm 7.0$ pb$^{-1}$ for 
this di-electron
data set~\cite{d0lepto1}.

In the paper of Ref.~\cite{d0lepto1} it is reported that no signal 
events exist in this
di-electron channel
where as the number of background events after applying all sets of cuts 
as described 
above is $0.44 \pm 0.06$~\cite{d0lepto1}. The uncertainty in background 
estimation is mainly due to the systematics. 
We exploit this information to obtain the limits of top squark pair 
production 
cross sections at 95\% C.L using Bayesian method for different 
choices of $\MST$ and 
for a given value of $\epsilon$. 
In this cross section limit calculation we take into account 
the uncertainties in background estimation, in the 
luminosity measurements and the uncertainty in detection efficiency.

Following the strategy as outlined above, using the data which implies 
that no signal events
with a background $0.44 \pm 0.06$,  we obtain limits of top squark pair
production cross sections
at 95\% C.L.    
for two choices of $\epsilon=$0.5 and 1. In fig.2 we show these limits
(solid lines) along with the theoretical predictions(dashed line). 
In theoretical calculation we multiply the K-factor
1.3 with the Born level cross section to take into 
account the next to leading order effect~\cite{spira}. 
Notice that, one can rule out
top squark mass upto 220(165) GeV for the choice of $\epsilon$=1(0.5) 
in a model independent way.
Moreover, as we explained already that the upper limits of  
cross sections which are consistent with di-electron data predict  
upper limits of
$\epsilon$ for a given $\MST$. 
In fig.3,  we present these upper limits 
of $\epsilon$ 
for each value of $\MST$ which is ruled out by data at 95\% C.L.
In the context of RPV SUSY model this 
upper limit of $\epsilon$ 
can be translated to obtain an upper limit of 
respective RPV couplings for a given
$\MST$ and SUSY parameter space. In fig.4 we show
the excluded region in the $\RPVC_{13j} -
\MST$ plane for a given set 
of SUSY parameters and for two choices of $\tan\beta=$5 and 30. 
In each figure we excluded region for two extreme values of 
$\cos\theta_{\tilde t}$=0.02 and 0.95.
The choice of our SUSY parameters for fig.3 and 4 
are(units are in GeV): 
\br
M_2&=&130,\ \  \mu=500,\ \  \tan\beta=5 (30). \nonumber \\
m_{\tilde\chi_{1}^\pm}&=& 515 (514),\ \ m_{\tilde\chi_{2}^\pm} 
= 121 (126), \nonumber \\
m_{\tilde\chi_{1}^0}&=& 63(65), 
\ \ m_{\tilde\chi_{2}^0}= 122(126) \nonumber \\
m_{\tilde\chi_{3}^0}&=& 504(506) 
\ \ m_{\tilde\chi_{4}^0}= 515(511) \nonumber \\
m_{\tilde q}&=&300, \ \ m_{\tilde\ell}=200, \ \ A_{b ,\tau, t}=200
\label{param}
\er
We discuss results in the next section.    

\vspace{1.0cm}
\section*{IV.~Results and Discussion}

We have computed the signal cross section in the
di-electron plus di-jet channel due to the  
top squark pair production at Tevatron in the framework of RPV SUSY 
model.
In addition to the RPC decay modes, $\tilde t_1$ also decays via two body 
decay channel into lepton and quark due to the presence of
RPV couplings. 
The relative rates of this RPV decay mode are shown in fig.1 as 
contours of fixed 
value of its BR $\epsilon$ in the 
$\lambda'_{13j}$ - $\MST$ plane for a given set of SUSY parameter space.
In the lower region of $\MST$ values, in addition to the RPV decay mode
(eq.~\ref{rpvdk})
other available RPC decay modes are the loop induced decay channel, 
(eq.~\ref{loopdk}) 
and 4-body decay mode, 
(eq.~\ref{4body}) which are of the same order in perturbation theory i.e 
${\cal O}(\alpha^3)$. In this mass region as expected, the RPV decay mode 
will dominate over the other decay modes for most of the parameter space 
depending on the value 
of ${\lambda'}_{13j}$ and
$\cos\theta_{\tilde t}$. Therefore, in this region of $\MST$, 
even very small value ($\sim 10^{-4}$)
of $\lambda'_{13j}$ coupling will yield appreciable rates for RPV 
decay
suppressing the two RPC decay modes. 
Once the value of $\MST$ crosses the $\MCH$(=112 GeV) threshold, 
the two body 
charged current decay mode (eq.\ref{chidk}) opens up, which is very 
much competitive to the RPV 
decay mode. As a result, large value
of RPV coupling $\lambda'_{13j}(\sim 0.1$) 
is needed to make the RPV decay 
mode comparable
with the two body charged current decay mode for a given $\MST$. 
We present this result 
in fig.1
for a single
SUSY parameter point. However, we have checked that this pattern 
more or less
exists  for an entire region of SUSY parameter space which are 
accessible at Tevatron.
We intend to emphasise that once the two body charged
current decay mode of $\tilde t_1$ opens up then it becomes dominant, 
otherwise the
RPV decay mode is the most dominant one in comparison to the loop level 
and 4-body decay modes.

In fig.2 we present the limiting values of signal cross section 
at 95\% C.L for 
two values of $\epsilon$(= 1 and 0.5).     
We also show the theoretical prediction of top 
squark pair production
cross section including K-factor~\cite{spira} by the dashed line 
in the same plane.
Comparing the cross section limits
with the theoretical predictions top squark masses  
can be constrained as a function of $\epsilon$. For instance, from the 
di-electron plus di-jet data we set the 
limit of top squark mass, $\MST \gsim $220(165) GeV for $\epsilon$=1(0.5). 
Notice that the limit of $\MST$ depends very strongly on 
$\epsilon$. However, there is a 10-30\% theoretical uncertainty due to 
the choice
of renormalisation and factorisation scales and parton distribution 
functions in cross section calculations. Note that these limits on $\MST$
are obtained in a model independent way.   

In fig.3, we show the upper limits of top squark decay 
BR($\epsilon$)
at 95\% C.L for various $\MST$ values.
These upper limits do not depend on any specific models. More precisely,
if $\tilde t_1$ state has the decay channel as eq.~\ref{rpvdk} then
the corresponding BR is restricted by existing data, as shown in fig.3.  
As for example, for $\MST$=100 GeV, 
the 95\% C.L. upper limit on $\epsilon$ is 0.35.

In the framework of RPV SUSY model the decay rate 
of $\tilde t_1$ is mainly controlled by $\lambda'_{13j}$ for a given
$\MST$ and $\cos\theta_{\tilde t}$(see eq.~\ref{rpvdk}). Therefore,
in this model, 
the upper limit of 
$\epsilon$ can be translated to the upper limit of $\lambda'_{13j}$
for a fixed $\MST$ and other SUSY parameters  
which determine the decay rates of $\t1$ into the  RPC decay modes
(see eq.~\ref{epsi}).
In fig.4, at 95\% C.L, we show the exclusion region 
in the $\lambda'_{13j} - \MST$
plane using data. The set of SUSY parameters
corresponding to this plot is given by eq.~\ref{param}.  
Notice that for a lower range of $\MST$ ($\sim 100$~GeV) where it 
appears to be NLSP, 
the RPV couplings are restricted to be
$\lambda'_{13j} \lsim 10^{-4}(10^{-3})$. 
Note that in this region for lower value of 
$\cos\theta_{\tilde t}$=0.02, the RPV decay rate is suppressed
(see eq.~\ref{rpvdk})
leading to weaker bounds where as for higher values of 
$\cos\theta_{\tilde t}$, bounds are turn out to be 
relatively better. However, the limits also become comparatively 
weaker   
in the higher side of $\MST$ where it is heavier than 
$\MCH$. 
This is because,
when the two body charged current decay mode of $\tilde t_1$ 
(eq.~\ref{chidk}) opens up,
then it becomes very competitive with the RPV decay mode  
leading lower BR for the RPV channel. Consequently, in this region, 
the BR limit constrains
only higher side of $\lambda'_{13j}$ for a given $\MST$.

It is obvious from eq.~\ref{rpvdk}, that the higher values of 
$\cos\theta_{\tilde t}$ will yield more stronger limits on 
$\lambda'_{13j}$. With the increase of $\tan\beta$,
the two body loop decay and 4-body decay  
~\cite{boehm, djouadi} widths enhances as the virtuality between
$\MST$ and $m_{\tilde\chi_{1,2}}$ decreases resulting a 
suppression of $\epsilon$ for a given $\MST$, which eventually leads a 
less constrained
region in the $\lambda'_{13j} - \MST$ plane.   
It is clear that for the region of $\MST \gsim \MCH$ i.e when $\tilde t_1$
is not NLSP, the bounds on RPV couplings are comparatively 
weaker.
           
\section*{V.~ Summary}

We investigate the di-electron plus di-jet signal due to the top 
squark pair 
production at Tevatron.
Identical final states also appear due to the 
first generation of Leptoquark production.
Exploiting the existing experimental data analysed by D0 group at 
Tevatron in the context of Leptoquark searches we try to constrain the top 
squark mass for various values of BR of top squarks in the RPV channel.
Using D0 data and assuming the BR of $\tilde t_1$ decay via the 
eq.~\ref{rpvdk},
$\epsilon$=1(0.5) we predict lower limits on $\MST \gsim$220(165)~GeV
at 95\% C.L. 
Repeating this exercise for various value of $\MST$
we exclude part of the parameter space in the $\epsilon - \MST$ plane 
as shown
in fig.3 in a model independent way. 

In the framework of RPV SUSY model, this exclusion region 
in $\epsilon - \MST$ plane  
converted 
to a corresponding exclusion region in the $\lambda'_{13j} - \MST$ plane as 
shown in fig.4 for a given set of SUSY parameter space.    
We exclude  
$\lambda'_{13j} \lsim 10^{-4}$ for 
$\MST=$100 GeV and 
$\tan\beta=5$ where as for high $\tan\beta$=30 region this limit
turns out to be relatively weak, $\lambda'_{13j} \lsim 10^{-3}$.    
Notice that when $\t1$ state appears to be NLSP,
the limits are very stringent and comparable to the limit,
for the case $\lambda'_{131}$, obtained from neutrino data~\cite{abada2} 
as discussed in Sec.I.
However, for, $\MST \gsim \MCH$, 
the present analysis does not give any better
limit than the others obtained from H1 experiments and APV measurements 
~\cite{H1,apv} and also from neutrino 
data~\cite{abada2}.  
We conclude that our predicted bounds are very stringent 
in the region where $\t1$ state appears to be NLSP.

\section*{Acknowledgement:}The authors are grateful to D. P. Roy for 
useful discussions and reading of the manuscript.    
MG is thankful to Sunanda Banerjee and Amitava Datta for many helpful 
discussions.

\newpage
%.........Inserting table..........................
\hspace{5 cm}
\vspace{1 cm}
\begin{center}
\begin{tabular}{|c|c|}
\hline
$m_{\tilde t_1}$(GeV) & Detection efficiency(\%) \\

&$e^{+}e^{-}$    \\
\hline
100 & 2.0 \\
120& 4.6 \\
140& 9.2\\
160&15.0\\
180&21.8\\
200&26.7\\
220&29.8\\
\hline
\end{tabular}
\end{center}

Table 1: Di-electron plus di-jet detection efficiencies for various  $\MST$.
%.........Inserting Figures..........................
%--------- Figure file- for Fig.1-----------------------

\begin{figure}[!b]
\vspace*{-3.5cm}
\hspace*{-3.0cm}
\mbox{\psfig{file=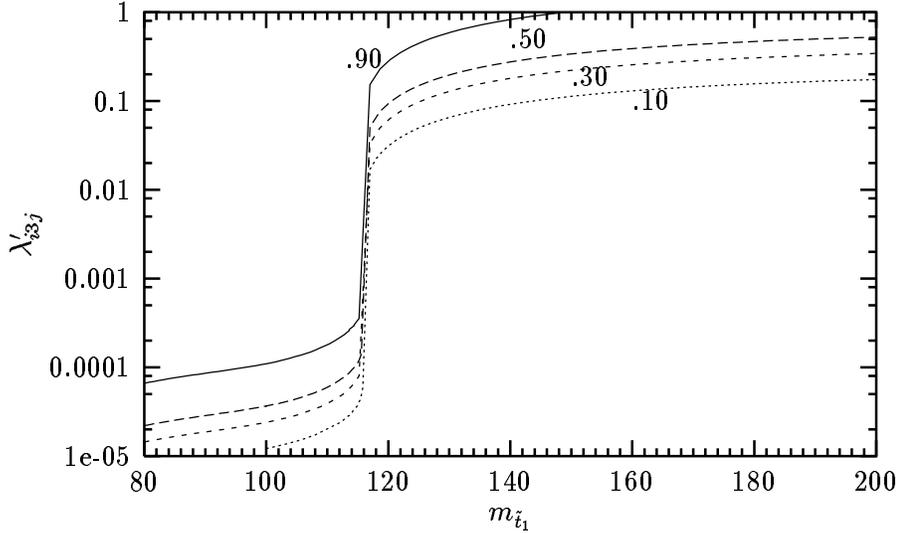,width=20cm}}
\vspace*{-16.7cm}
\caption{ Branching ratio contours for the decay channel
$\tilde t_1 \ra \ell + q$. 
The SUSY parameters are: 
$M_2=125$ GeV, $\mu=400$ GeV, $\tan\beta=$4, $\cos\theta_{\tilde t}=$0.8
and $m_{\tilde q}=300$ GeV, $m_{\tilde \ell}=200$ GeV, 
$A_b=A_\tau=200$ GeV.  
}
\label{fig_mssm1}
\end{figure}
\pagebreak
\begin{center}
\includegraphics[width=5.in, height=3.in]{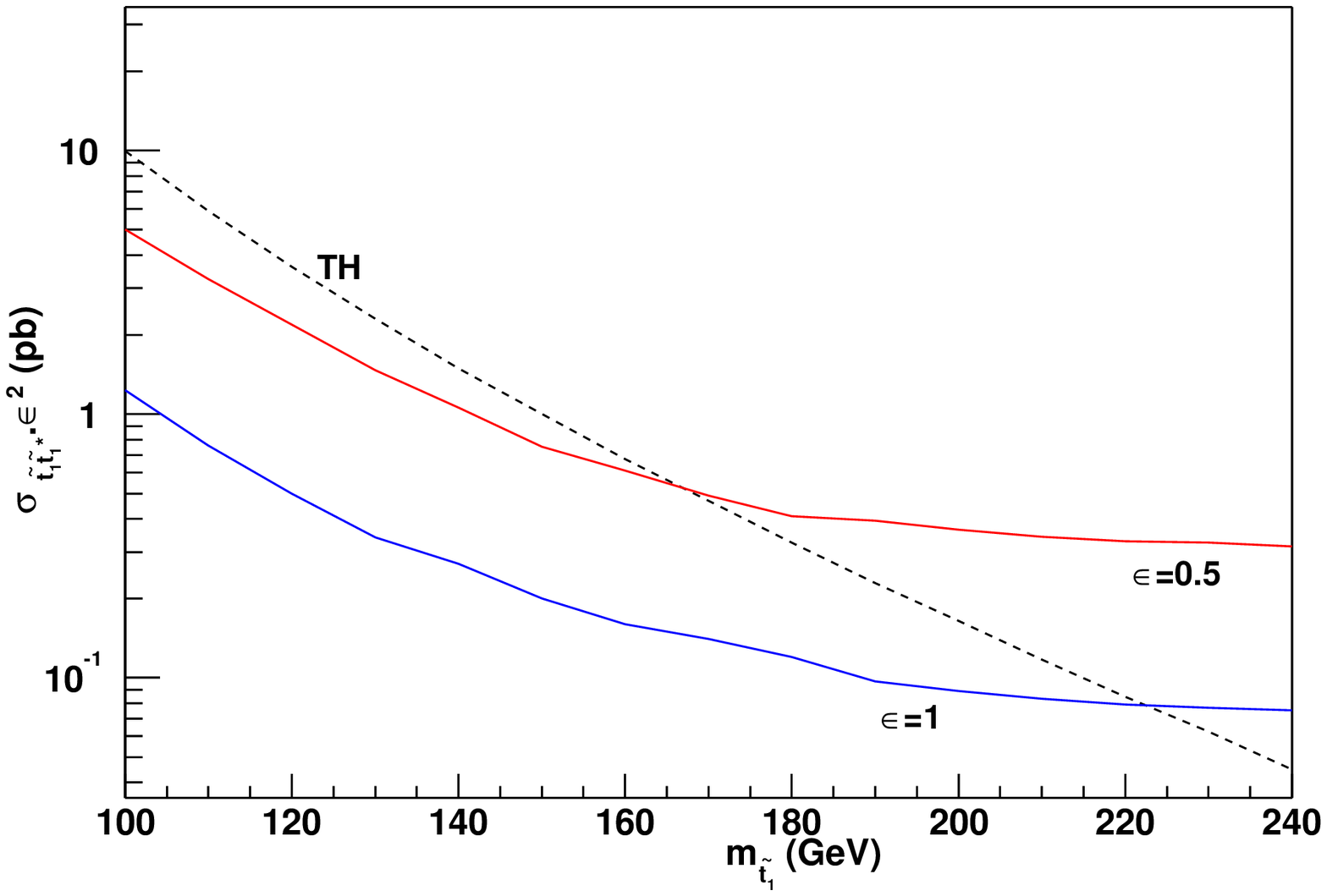}
\vspace{1.cm}
\\
\end{center}

Figure 2: The top squark pair production cross section limits  
at 95\% C.L (solid lines) for
$\epsilon$=1 and 0.5 along with the 
theoretical prediction(dashed line).    

\begin{center}
\includegraphics[width=5.in, height=3.in]{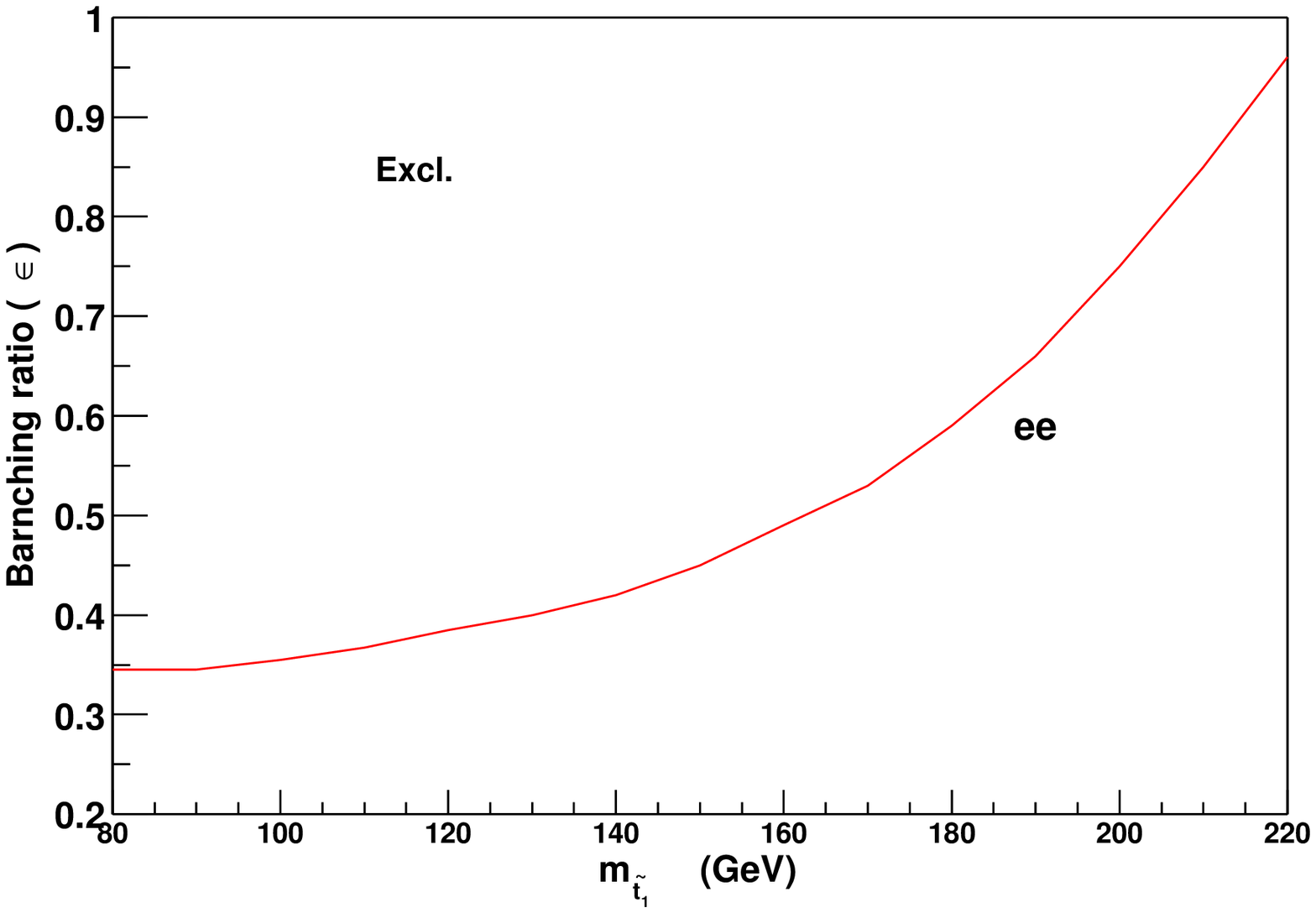}
\vspace{1.cm}
\\
\end{center}

Figure 3: The excluded region by di-electron data at 95\% C.L.

\begin{center}

\includegraphics[width=5.in, height=3.in]{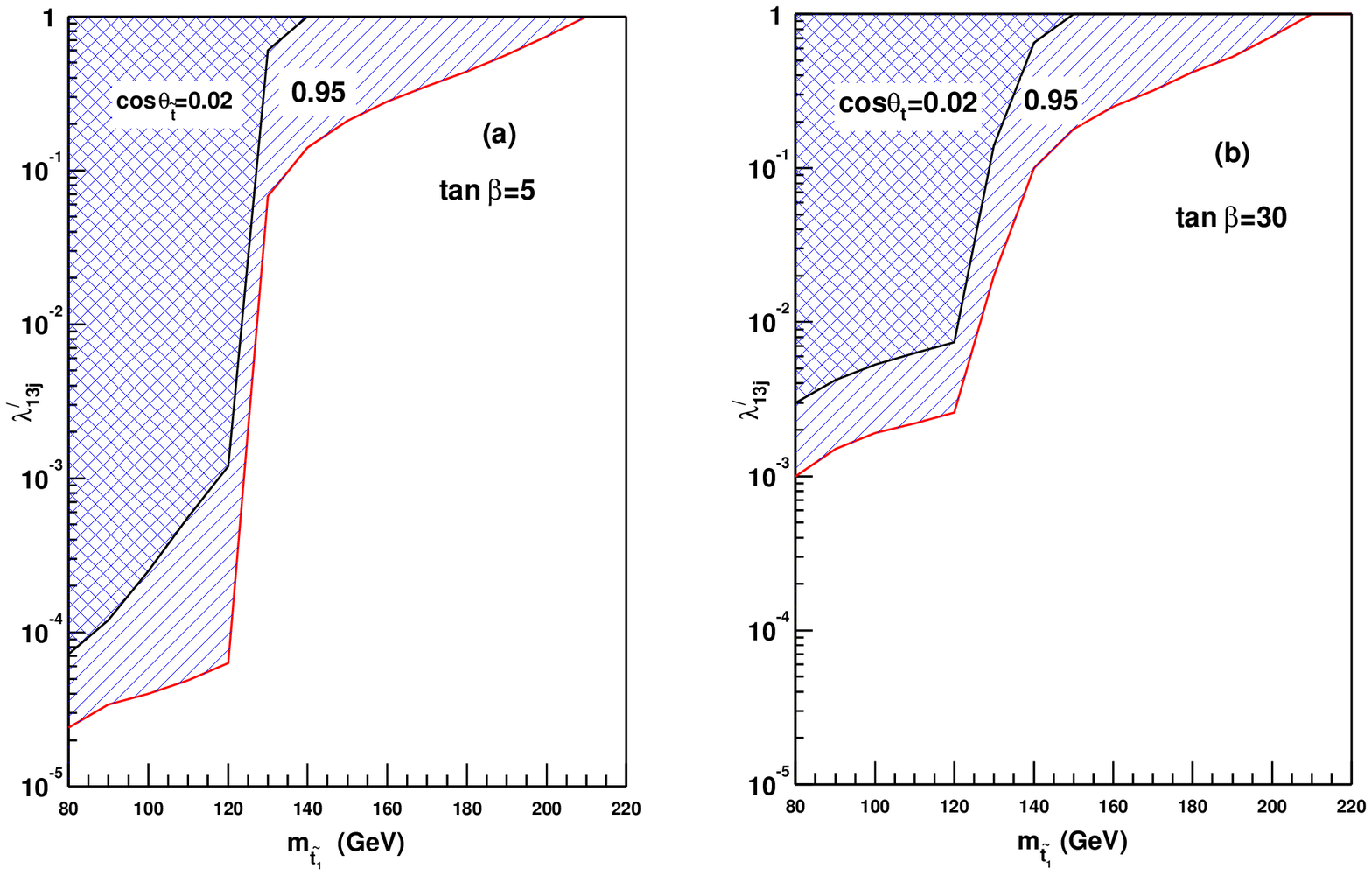}
\vspace{1.0cm}
\\
\end{center}
Figure 4: The excluded region(hatched) by di-electron data  
at 95\% C.L. 
The SUSY parameters are: $M_2=$130 GeV, $\mu$=500 GeV,
$m_{\tilde q}=$300 GeV, $m_{\tilde\ell}=$200 and A-terms=200 GeV.

\end{document}